\documentclass{iaus}
\usepackage{graphicx}
\usepackage{ifpdf}

\title{Building Blue Stragglers with Stellar Collisions}

\author{Evert Glebbeek \and Onno R.~Pols}
\affiliation{Sterrekundig Instituut Utrecht, P.O. Box 80000, 3508 TA Utrecht, The Netherlands}

\pubyear{2007}
\volume{246}  
\jname{Proceedings Dynamical Evolution of Dense Stellar Systems}
\editors{E. Vesperini, M. Giersz \& A. Sills, eds.}

\providecommand{\citep}[1]{(\cite{#1})}
\providecommand{\citet}[1]{\cite{#1}}

\begin{document}

\maketitle

\begin{abstract}
The evolution of stellar collision products in cluster simulations has
usually been modelled using simplified prescriptions. Such prescriptions either
replace the collision product with an (evolved) main sequence star, or assume
that the collision product was completely mixed during the collision.

It is known from hydrodynamical simulations of stellar collisions that
collision products are not completely mixed, however. We have calculated
the evolution of stellar collision products and find that they are brighter
than normal main sequence stars of the same mass, but not as blue as models
that assume that the collision product was fully mixed during the
collision.
\end{abstract}

\section{Introduction}
The aim of the MODEST collaboration \citep{article:modest1} is to model and
understand dense stellar systems,
which requires a good understanding of what happens when two single stars or binary
systems undergo a close encounter. A possible outcome of such an encounter is a
collision followed by the merging of two or more stars. This is a possible formation
channel for blue straggler stars (\emph{e.g.} \citet{article:sills_on_axis}).
Understanding the formation and evolution of blue stragglers is important for
understanding the Hertzsprung-Russell diagram of clusters.

\section{Method}
We have developed a version of the Eggleton stellar evolution code
\citep{article:eggleton_evlowmass, article:pols_approxphys}
that can import the output of a collision calculation and calculate the 
subsequent evolution of the remnant, in principle without human intervention. 

We have used this code to calculate detailed evolution models of collision 
remnants from the $N$-body simulation of M67 by \citet{article:hurley_m67}
and compared these with normal main sequence stars of the same mass as well as
homogeneous models with the same average abundances. The post-merger profiles 
were calculated with the parametrised code of \citet{article:lombardi_mmas}.

The collisions in the $N$-body simulation span a range of total masses 
$M = M_1+M_2$, mass ratio $q=M_2/M_1$ and time of collision $t$. We have
also calculated a grid of models spanning
four collision times ($t = 2800, 3100, 3400 \mbox{ and } 3700 \mathrm{M yr}$), 
four mass ratios ($q =  0.4, 0.6, 0.8 \mbox{ and } 1.0$) 
and six total masses ($M = 1.5, 1.6, 1.7, 1.8, 1.9 \mbox{ and } 2.0$), which
covers the parameter space of collisions found in the $N$-body simulation.

\section{Results and Conclusions}

\begin{figure}
\ifpdf
   \includegraphics[width=\textwidth]{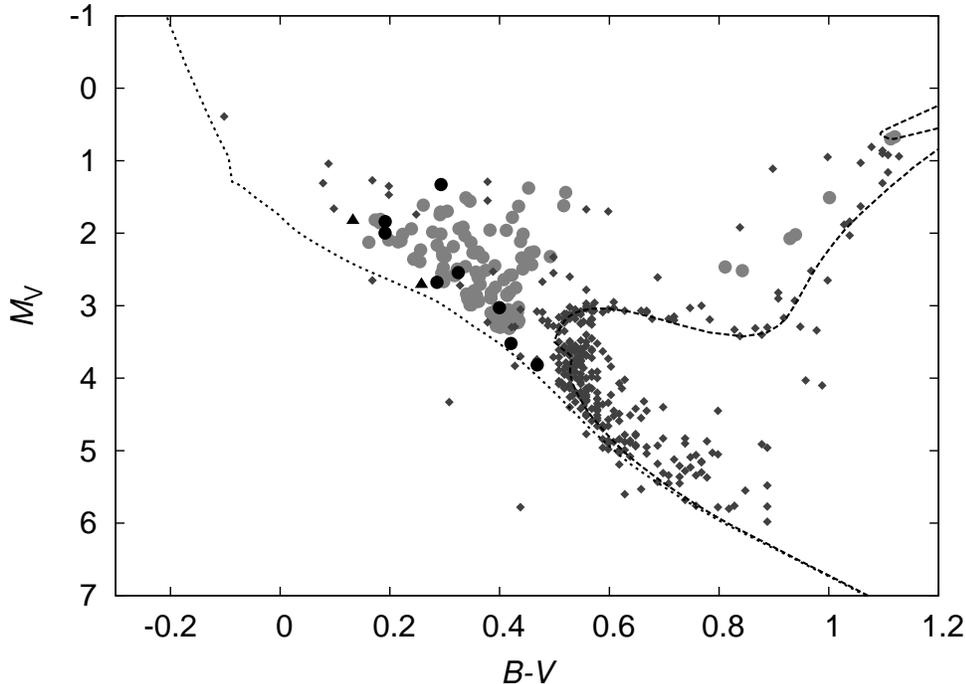}
\else
   \includegraphics[angle=270,width=\textwidth]{glebbeek_fig_hrd}
\fi
\caption{Colour-magnitude diagram of the open cluster M67 
($\blacklozenge$). Overplotted are the locations of our models at 
$4 \mathrm{Gyr}$, the age of M67. The black 
({\Large$\bullet$, $\blacktriangle$}) symbols are collisions from the M67 
simulation. Two of these are double collisions, which are indicated by
{\Large$\blacktriangle$}. The grey ({\Large$\bullet$}) symbols are from our 
larger grid.}
\label{fig:hrd_m67}
\end{figure}

Compared to normal stars, collision products are helium enhanced. Most of the 
helium enhancement is in the interior and does not affect the opacity of the
envelope. 
The helium enhancement does increase the mean molecular weight and therefore
the luminosity of the star. This decreases the remaining lifetime of collision 
products compared to normal stars of the same mass and can be important for
the predicted number of blue stragglers from cluster simulations. The increased
luminosity changes the distribution of blue stragglers in the colour-magnitude
diagram, moving it above the extension of the main sequence.

The evolution track of a fully mixed model can be significantly bluer than
a self-consistently calculated evolution track of a merger remnant. Fully
mixed models are closer to the zero age main sequence.

Our grid of models covers most of the observed blue straggler region of M67 
(Figure \ref{fig:hrd_m67}). A better coverage of the blue part of the region
can be achieved by increasing the upper mass limit in the grid. The brightest
observed blue straggler falls outside the region of our grid because it
requires at least a double collision to explain.

\bibliography{glebbeek_bibliography}
\end{document}